\documentclass[aps,prl,twocolumn,nofootinbib,
superscriptaddress,showpacs,preprintnumbers,10pt]{revtex4-1}

\usepackage{amsmath,amsfonts,amssymb}
\usepackage{graphicx,bm,color,subfigure,hyperref,xspace}

\DeclareMathOperator{\im}{Im}
\newcommand{\diff}{\ensuremath{\mathrm{d}}}
\newcommand{\ie}{{\it i.e.}\xspace}
\newcommand{\rhs}{{\it rhs}\xspace}
\newcommand{\wbi}{\ensuremath{\widehat \beta_i(t)}\xspace}
\newcommand{\gev}{\ensuremath{\,\text{Ge\kern -0.1 em V}}\xspace}
\newcommand{\etamaid}{{$\eta$-MAID}\xspace}

\begin{document}

\title{Analyticity Constraints for Hadron Amplitudes: \\
Going High to Heal Low Energy Issues}

\author{V.~Mathieu}
\email{mathieuv@indiana.edu}
\affiliation{Theory Center, Thomas Jefferson National Accelerator Facility,
Newport News, VA 23606, USA}
\affiliation{Center for Exploration of Energy and Matter, 
Indiana University, Bloomington, IN 47403, USA}
\affiliation{Physics Department, Indiana University, Bloomington, IN 47405, USA}

\author{J.~Nys}
\email{jannes.nys@ugent.be}
\affiliation{Center for Exploration of Energy and Matter, 
Indiana University, Bloomington, IN 47403, USA}
\affiliation{Department of Physics and Astronomy, Ghent University, Belgium}

\author{A.~Pilloni}
\affiliation{Theory Center, Thomas Jefferson National Accelerator Facility,
Newport News, VA 23606, USA}

\author{C.~Fern\'andez-Ram\'irez}
\affiliation{Instituto de Ciencias Nucleares, 
Universidad Nacional Aut\'onoma de M\'exico, Ciudad de M\'exico 04510, Mexico}

\author{A.~Jackura}
\affiliation{Center for Exploration of Energy and Matter, 
Indiana University, Bloomington, IN 47403, USA}
\affiliation{Physics Department, Indiana University, Bloomington, IN 47405, USA}

\author{M.~Mikhasenko}
\affiliation{{Universit\"at Bonn, 
Helmholtz-Institut f\"ur Strahlen- und Kernphysik, 53115 Bonn, Germany}}

\author{V.~Pauk}
\affiliation{Theory Center, Thomas Jefferson National Accelerator Facility,
Newport News, VA 23606, USA}

\author{A.~P.~Szczepaniak}
\affiliation{Theory Center, Thomas Jefferson National Accelerator Facility,
Newport News, VA 23606, USA}
\affiliation{Center for Exploration of Energy and Matter, 
Indiana University, Bloomington, IN 47403, USA}
\affiliation{Physics Department, Indiana University, Bloomington, IN 47405, USA}

\author{G. Fox}
\affiliation{School of Informatics, Computing, and Engineering, 
Indiana University, Bloomington, IN 47405, USA} 

\collaboration{Joint Physics Analysis Center}
\noaffiliation
\preprint{JLAB-THY-17-2539}

\begin{abstract}
Analyticity constitutes a rigid constraint 
on hadron scattering amplitudes.
This property is used to relate models 
in different energy regimes.
Using meson photoproduction as a benchmark, 
we show how to test contemporary 
low energy models directly against high energy data. 
This method pinpoints deficiencies of the models 
and treads a path to further improvement.
The implementation of this technique 
enables one to produce more stable and reliable partial waves 
for future use in hadron spectroscopy and new physics searches.
\end{abstract}
\maketitle

\textit{Introduction.}--- Determination of various 
hadronic effects represents a major challenge 
in searches for New Physics through precision
measurements~\cite{Archilli:2017nbp,Ciezarek:2017nbp,pbf,Jegerlehner:2009ry,Cirigliano:2013lpa}.
For example, the possible identification of 
Beyond Standard Model signals in $B$ meson decays is hindered 
by uncertainties in hadronic final state interactions.
The strongly coupled nature of QCD prevents us 
from computing these effects directly 
from the underlying microscopic formulation. 
Nevertheless, one can use the first principles of 
$S$-matrix theory to impose stringent constraints 
on hadron scattering 
amplitudes~\cite{smatrix,Chew:1961,Chew:1972na}. 
These approaches are encountering a renewed interest
even in the more formal context of strongly coupled
theories~\cite{Paulos:2016fap,Paulos:2016but,Paulos:2017but}.

In this Letter, we show how to use analyticity 
to relate the amplitudes at high energies to the physics 
at low energies, where resonance effects dominate.
This is not only important for reducing hadronic 
uncertainties in the aforementioned processes, 
but is of interest on its own merits for 
unraveling the spectrum of QCD.
According to  
phenomenological predictions and lattice QCD simulations, the current spectrum summarized in the 
Particle Data Group (PDG) is far from 
complete~\cite{pdg}.
For example, the recent discoveries of 
unexpected peaks in data indicate that 
the true hadron spectrum is far more complex than
predicted~\cite{Pennington:2016dpj,Briceno:2015rlt,Aaij:2015tga,Meyer:2015eta,Lebed:2016hpi,Esposito:2016noz}.
As a working case, 
we focus here on the baryon sector 
in the intermediate energy range. 
In the PDG these $N^*$ and $\Delta$ resonances are referred to 
as ``poorly known"~\cite{pdg}, 
despite the large amount of data available. 
The ambiguities encountered when identifying resonances 
are related to the fact that, 
as the center of mass energy increases, 
so does the number of contributing partial waves, 
vastly complicating the reaction models used in data analysis.
The $2-3\gev$ mass region is of particular interest 
for baryon spectroscopy since, 
besides the ordinary quark model 
multiplets, it is expected to contain 
a new form of exotic light quark matter 
that is dominated by excitations of the gluon
field~\cite{Dudek:2012ag,Meyer:2015eta}.
The recent upgrade at Jefferson 
Lab~\cite{battaglieri2011meson,battaglieri2016baryon,Dudek:2012vr,AlGhoul:2017nbp,Shepherd:2009zz} 
is providing high statistics data on hadron photoproduction.
New amplitude analysis methods are a prerequisite 
to achieve a robust extraction of hadron resonance parameters.

Many research groups carry out low energy, coupled channel,
partial wave analyses (PWA) for baryon spectroscopy. 
Currently, the most active are 
ANL-Osaka~\cite{Kamano:2013iva},
Bonn-Gatchina~\cite{Anisovich:2011fc,Anisovich:2012ct},
JPAC~\cite{Fernandez-Ramirez:2015tfa,Fernandez-Ramirez:2015fbq}, 
J\"ulich-Bonn~\cite{Ronchen:2014cna,Ronchen:2015vfa}, 
MAID~\cite{Chiang:2001as},  
and SAID~\cite{Workman:2012jf,Briscoe:2012ni}. 
These groups perform global fits to 
hadro- and/or photoproduction data 
using a finite set of partial waves 
to extract baryon resonance
properties~\cite{Beck:2016hcy,Ruic:2011wf}. 
In these approaches the high energy data 
are largely ignored. 
As we show in this Letter,
these data can greatly impact the baryon spectrum analyses 
through analyticity. 
Specifically, we implement Finite Energy Sum Rules 
derived from dispersion relations~\cite{Dolen:1967zz}, 
and use simple approximations to describe the high energy data.
The sum rules relate the amplitudes in the baryon resonance region 
to the high energy dynamics, where the amplitudes are described by exchanges of meson Regge poles ~\cite{Collins:1977jy}.
We apply our method to the existing data on 
$\pi^0$ and $\eta$
photoproduction~\cite{Braunschweig:1970jb,Anderson:1971xh,Dewire:1972kk}.
These cases constitute a first step towards a  
straightforward and systematic implementation 
of high energy constraints into low energy amplitudes, 
and provide a template for further application in data analysis. 

\begin{figure}[t]
\centering
\includegraphics[width=0.8\linewidth]{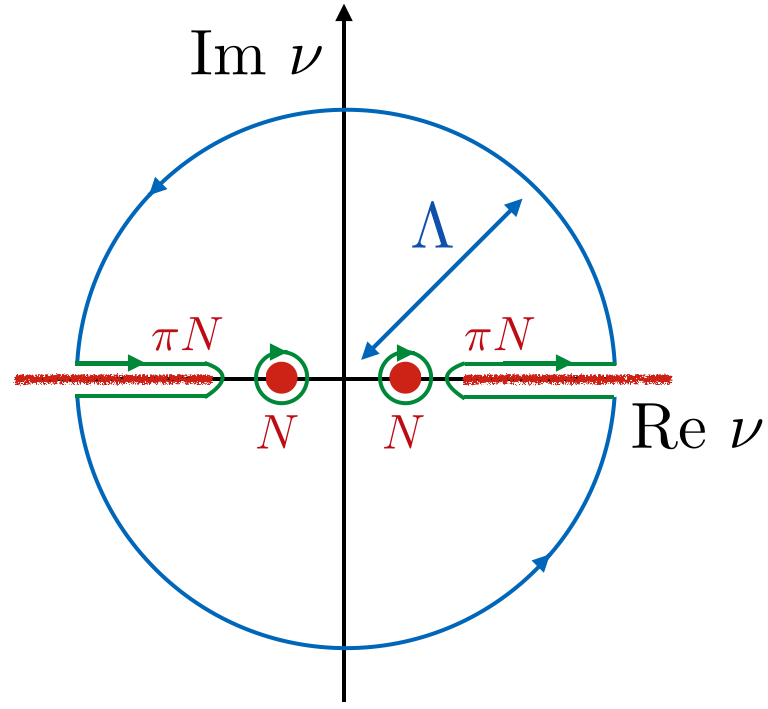}
\caption{Contour in the complex $\nu$-plane used in the derivation 
of the sum rules in Eq.~\eqref{eq:FESR}. 
The radius $\Lambda$ must be taken sufficiently large, 
for the single Regge pole approximation to hold at $|\nu| = 
\Lambda$.
The nucleon pole and  the $\pi N$ cuts 
are shown  on the real axis. }\label{fig:nuplane}
\end{figure}

\textit{Analyticity constraints for photoproduction.}--- 
The reaction $\gamma p \to x p$, 
where $x=\pi^0,\eta$ is completely described in terms of 
four independent scalar amplitudes $A_i(s,t)$. 
These are analytic functions of the 
Mandelstam variables $s$ (the square of the center of mass energy) 
and $t$ (the square of the momentum transfer)~\cite{Chew:1957tf}. 
At fixed $t$, each $A_i(s,t)$ satisfies an unsubtracted dispersion relation 
involving the discontinuity with respect to $s$ 
along the unitarity cut 
and the crossed-channel unitarity cut 
in $u = 2m_p^2 + m_x^2 - s -t$.
Charge conjugation symmetry relates the discontinuity along 
the crossed channel cut to that of the direct channel. 
This symmetry is made explicit 
by writing the amplitude as a function of the 
variable\footnote{As customary, 
all dimensional variables are given in units of $1\gev$.} 
$\nu \equiv (s-u)/2$. For large $|\nu|$ and small 
$t$ kinematics, 
the amplitudes are well approximated by Regge poles, 
\ie via crossed channel exchanges. 
In this region, the amplitudes take the form  
\begin{equation}
\im A_i(\nu,t) = \sum_n \beta^{(n)}_i(t) \,\nu^{\alpha^{(n)}(t)-1}. 
\label{eq:regge}
\end{equation}
The Regge poles are determined 
by the trajectories $\alpha^{(n)}(t)$ 
and the residues $\beta^{(n)}_i(t)$.
The index $n$ runs over all possible exchanges.
This approximation
holds only if $|\nu|$ is greater than some cutoff 
$\Lambda$ above 
the resonance region. For $|\nu| < \Lambda$, 
the amplitude is dominated by direct channel resonances, 
and thus it can be well approximated 
by a finite number of partial waves. 
One can write a dispersion relation using  Cauchy's theorem with the contour 
in Fig.~\ref{fig:nuplane}, 
and calculate explicitly the integral in the circle 
$|\nu| = \Lambda$ assuming the form in Eq.~\eqref{eq:regge}. One readily obtains~\cite{Collins:1977jy} 
\begin{align} \label{eq:FESR}
\int_0^\Lambda \im A_i(\nu,t)\ \nu ^k \: \diff \nu & =
\sum_{n} \beta_i^{(n)}(t) \: \frac{\Lambda^{\alpha^{(n)}(t)+k}}{\alpha^{(n)}(t)+k}.
\end{align}
The amplitudes $A_{1,2,4}$ and $A_3$ 
are even and odd functions of $\nu$, respectively. 
Here $k$ is an arbitrary positive integer, 
odd for $A_{1,2,4}$ and even for  $A_3$. 
We give the value of $\Lambda$ in terms of an 
energy cutoff $s_\text{max}$, 
which introduces additional $t$ dependence 
$\Lambda = s_\text{max} + (t-2 m_p^2-m_x^2)/2$. 
We restrict the sum on the right hand side (\rhs) 
of Eq.~\eqref{eq:FESR} 
to the dominant $t$-channel Regge poles. 
Each $A_i$ receives a   contribution from both 
isoscalar and isovector exchanges. 
Natural parity exchanges (with $P = (-)^J$) 
dominate $A_1$ and $A_4$, 
while the unnatural ones (with $P = (-)^{J+1}$) 
dominate $A_2$ and $A_3$. 
More specifically, the  $n=\rho, \omega$, Regge poles contribute 
to $A_1$ and $A_4$, while $A_2$ and $A_3$ are determined 
by exchanges of the 
$n=b,h,\rho_2,\omega_2$.\footnote{Even though there are 
some experimental indications 
of the existence of $\rho_2$ and
$\omega_2$~\cite{Anisovich:2002su,Anisovich:2011sva}, 
they have been observed by one single group, 
and thus need further confirmation~\cite{pdg}.}

\begin{figure*}[t]
\centering
\subfigure[~\label{fig:Res-Pi0-1}]{\includegraphics[width=.24\linewidth]{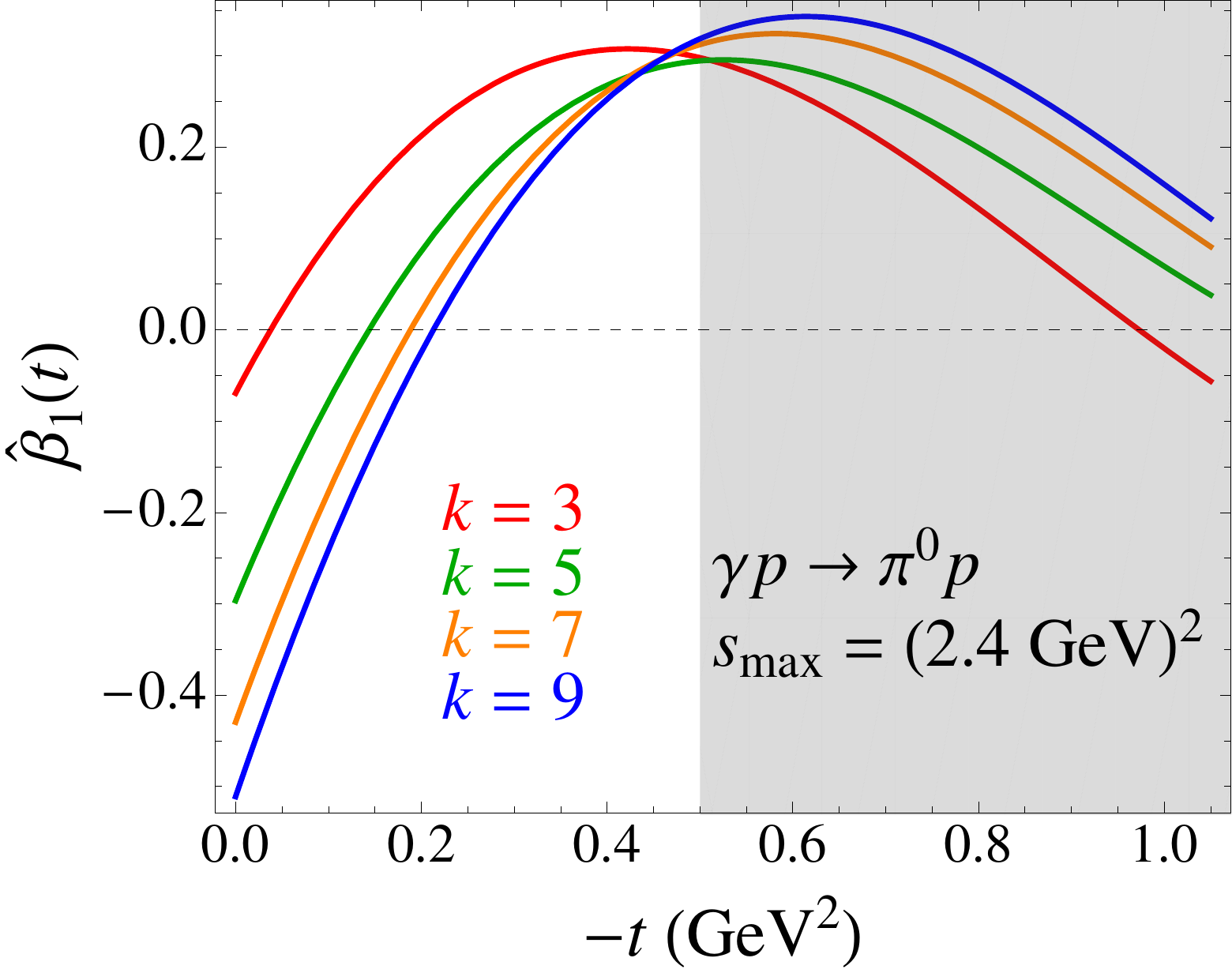}}
\subfigure[~\label{fig:Res-Pi0-4}]{\includegraphics[width=.24\linewidth]{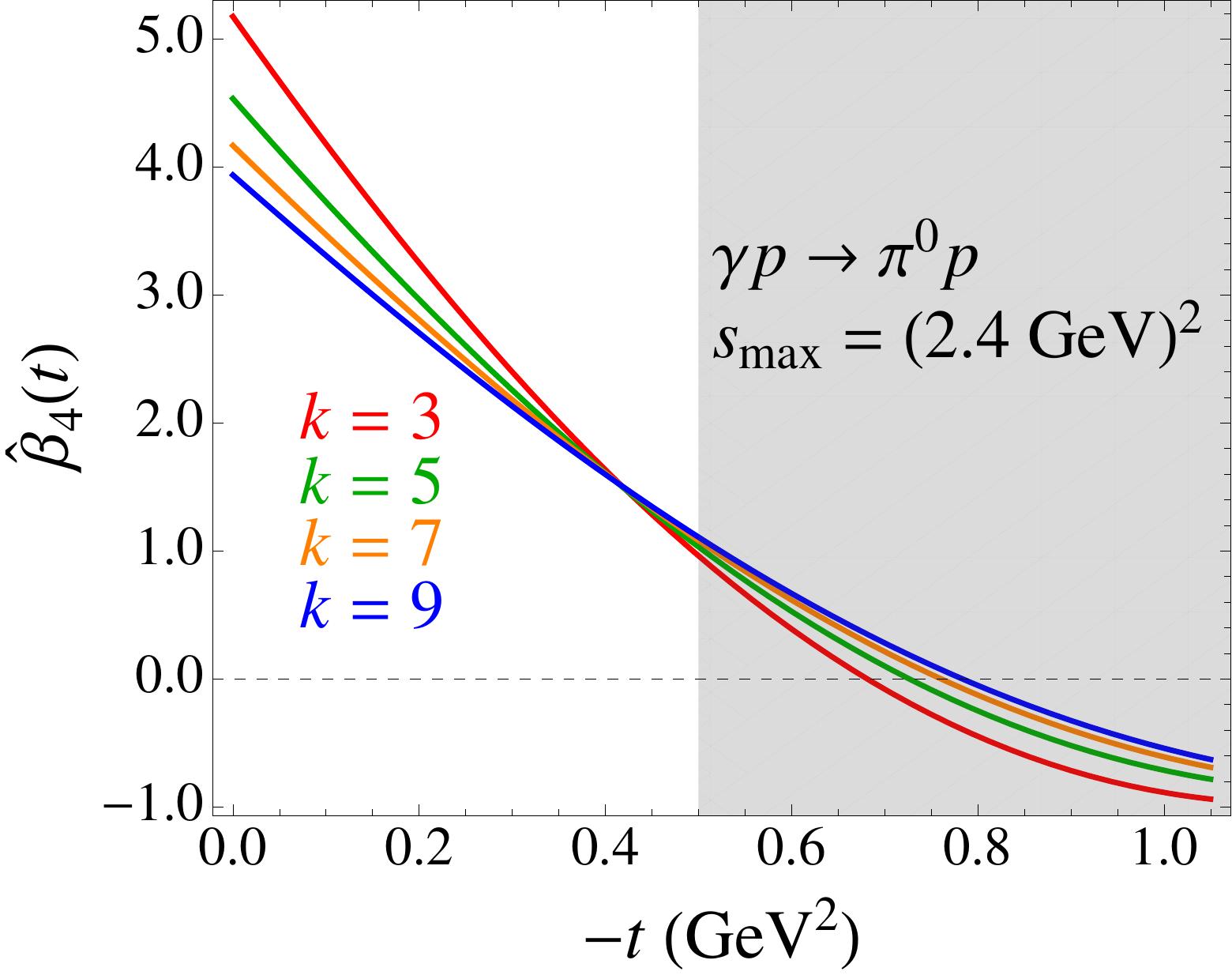}}
\subfigure[~\label{fig:Res-Eta-1}]{\includegraphics[width=.24\linewidth]{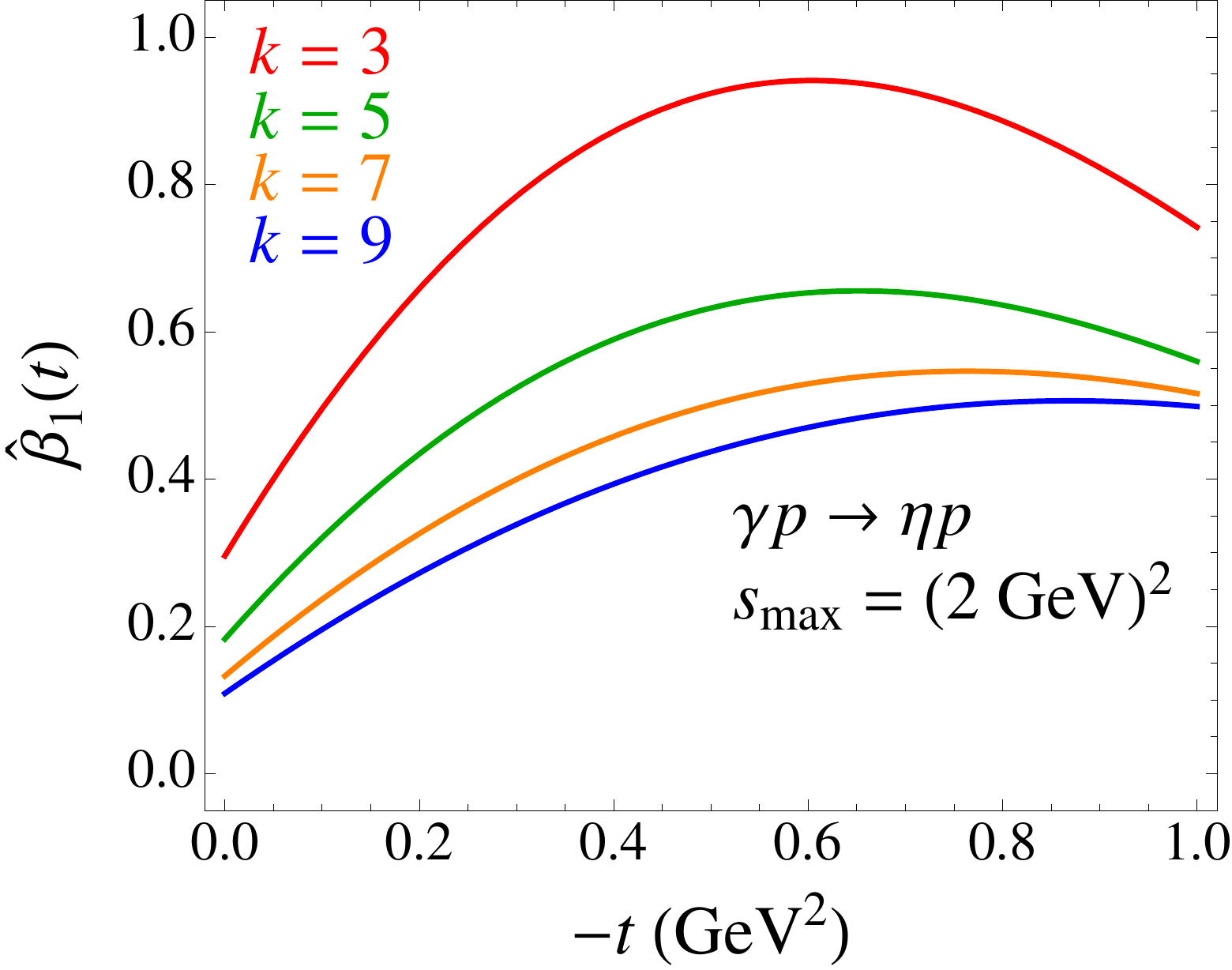}}
\subfigure[~\label{fig:Res-Eta-4}]{\includegraphics[width=.24\linewidth]{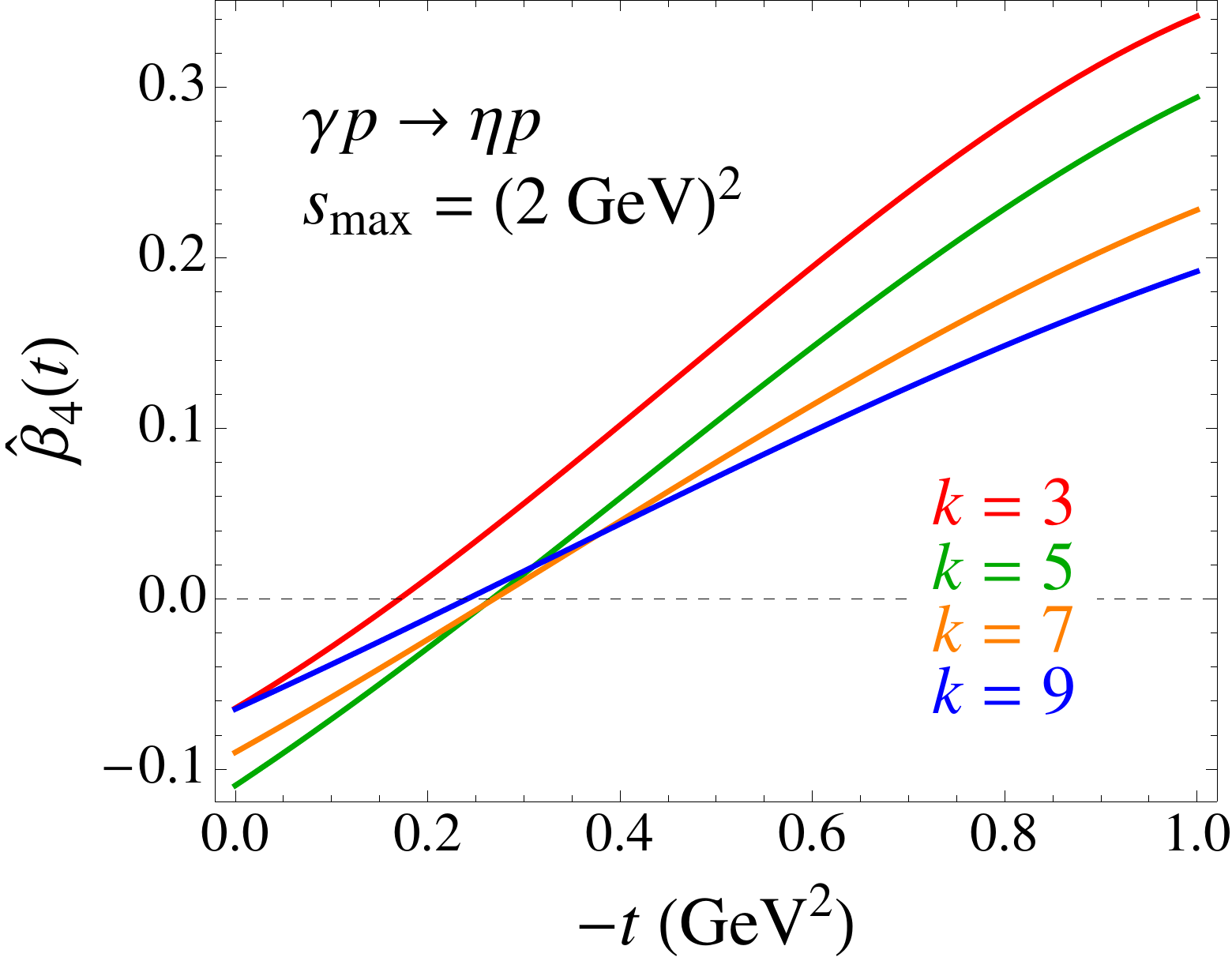}}
\caption{\label{fig:wbi} Effective residues computed 
from the low energy models using 
Eq.~\eqref{eq:betaRHS}. 
(a) and (b): $\pi^0$ photoproduction 
using SAID~\cite{Workman:2012jf}.
(c) and (d): $\eta$ photoproduction 
using \etamaid~\cite{Chiang:2001as}. 
For $\pi^0$, the single pole approximation 
is valid for $- t \lesssim 0.5 ~\gev^2$, 
as explained in the text.  
The dispersion in $k$ is small for $\pi^0$, 
while the large variation with $k$ for $\eta$ 
indicates  issues with the low energy model.
}
\end{figure*}

The trajectories are nearly degenerate for all the 
natural exchanges~\cite{Mandula:1970wz}, 
and in the kinematical region of interest, they can be 
well approximated by 
\begin{equation}
\alpha(t) \equiv \alpha^{(\rho)}(t) = 
\alpha^{(\omega)}(t) = 1+ 0.9 \, (t-m_\rho^2),\label{eq:alpha1}
\end{equation}
for $i=1,4$. 
For the unnatural exchanges, 
\mbox{$\alpha(t) =  0.7 \, (t-m_\pi^2)$} for $i=2,3$.
At high energy the contribution of unnatural versus natural exchanges 
to observables  
in the forward direction is suppressed.
For example, with a beam energy of $9\gev$, 
the suppression is expected to be 
\mbox{$\nu^{2(0.9 m_\rho^2-0.7m_\pi^2-1)} \sim 7\%$}. 
This can be compared with polarization observables, 
such as the beam asymmetry
$\Sigma$,\footnote{The beam asymmetry is 
$\Sigma \equiv (\diff \sigma_\perp - \diff \sigma_\parallel)/
(\diff \sigma_\perp + \diff \sigma_\parallel)$, 
with $\diff \sigma_{\perp(\parallel)}$ 
the differential cross section of the photon polarized 
perpendicular (parallel) 
to the reaction plane.} 
which are sensitive to the interference between 
the natural and the unnatural Regge poles. 
If one neglects the unnatural contributions, $\Sigma = 1$.
The recent measurement of $\pi^0$ and $\eta$ beam asymmetries at 
GlueX~\cite{AlGhoul:2017nbp} confirms that $\Sigma  > 0.9 $, 
so that the unnatural exchanges contribute $\lesssim 5\%$ 
to the observables.
In the following, we will consider the amplitudes 
dominated by natural exchanges, 
$A_1$ and $A_4$, only. 
We use low energy models as input to determine 
the left hand side of Eq.~\eqref{eq:FESR}, 
and use it to predict the residues.
To this aim we define the effective residues,
\begin{align} \label{eq:betaRHS}
\widehat \beta_i(t) & = \frac{\alpha(t)+k}{\Lambda^{\alpha(t)+k}}
\int_0^\Lambda \im A^\text{PWA}_i(\nu,t)\, \nu^k \: \diff \nu ,
\end{align}
where $A^\text{PWA}_i$ is the amplitude calculated 
from low-energy models.

Because of Regge trajectory degeneracy, 
the \wbi's describe the sum of the contribution 
of both isovector and isoscalar exchanges. 
Consistency of the single pole hypothesis requires  
the \rhs of Eq.~\eqref{eq:betaRHS} 
to be independent of $k$.\footnote{For example, 
if one added another nondegenerate trajectory 
$\alpha_{2} < \alpha$, 
the effective residue would depend on $k$ as  
$ \hat\beta_i = \beta_i 
+ \frac{\beta_{i,2}}{\Lambda^{\alpha-\alpha_{2}}}
\frac{\alpha+k}{\alpha_{2}+k}$. 
The latter becomes negligible for $\Lambda$ sufficiently large.} 
For $|\nu| > \Lambda$, the amplitudes 
can be expressed in terms of the effective residues as~\cite{Collins:1977jy}
\begin{equation}\label{eq:Airec} 
\widehat A_i(\nu,t) = 
\left[i+ \tan\frac{\pi}{2} \alpha(t) \right] 
\widehat \beta_i(t)\, \nu^{\alpha(t)-1}.
\end{equation}
The $\widehat A_i(\nu, t)$ 
are the high energy amplitudes calculated 
from the low energy models entering in the \wbi. 
Comparing the observables calculated with these 
to data allows us to check the quality of the low energy models.
In the high energy limit, the differential cross section becomes
\begin{align}
\frac{\diff  \widehat \sigma}{ \diff t} &\simeq \frac{1}{32 \pi} 
\left[ \left |\widehat A_1 \right|^2 
- t \left |\widehat A_4\right|^2 \right] \nonumber\\
& =  \frac{\nu^{2\alpha(t)-2} }{32\pi}  
\left[1+ \tan^2 \frac{\pi}{2}\alpha(t) \right] 
\left[ \widehat\beta_1^2(t) - t \: \widehat\beta^2_4(t) \right].
\label{eq:dsdt-bis}
\end{align}

\textit{Results.}--- We next discuss what these constraints can 
tell us about the existing low energy analyses.
We consider \wbi for $k=3,5,7,9$.
For $\pi^0$, we use the SAID partial wave model 
which is valid up to  
$s_{\text{max}}= (2.4\gev)^2$~\cite{Workman:2012jf}.  
For $\eta$, the amplitudes need to be extrapolated 
below the physical $\eta N$ threshold, 
down to the $\pi N$ threshold (see Fig.~\ref{fig:nuplane}).  
Among the various models, only \etamaid~\cite{Chiang:2001as}, 
valid only up to $s_{\text{max}}= (2\gev)^2$, 
is given in terms of analytical functions  
that allow for this continuation~\cite{Nys:2016vjz}. 

\begin{figure*}[htb]
\begin{center}
\subfigure[~\label{fig:Sig-Pi0}]{\includegraphics[width=0.85\columnwidth]{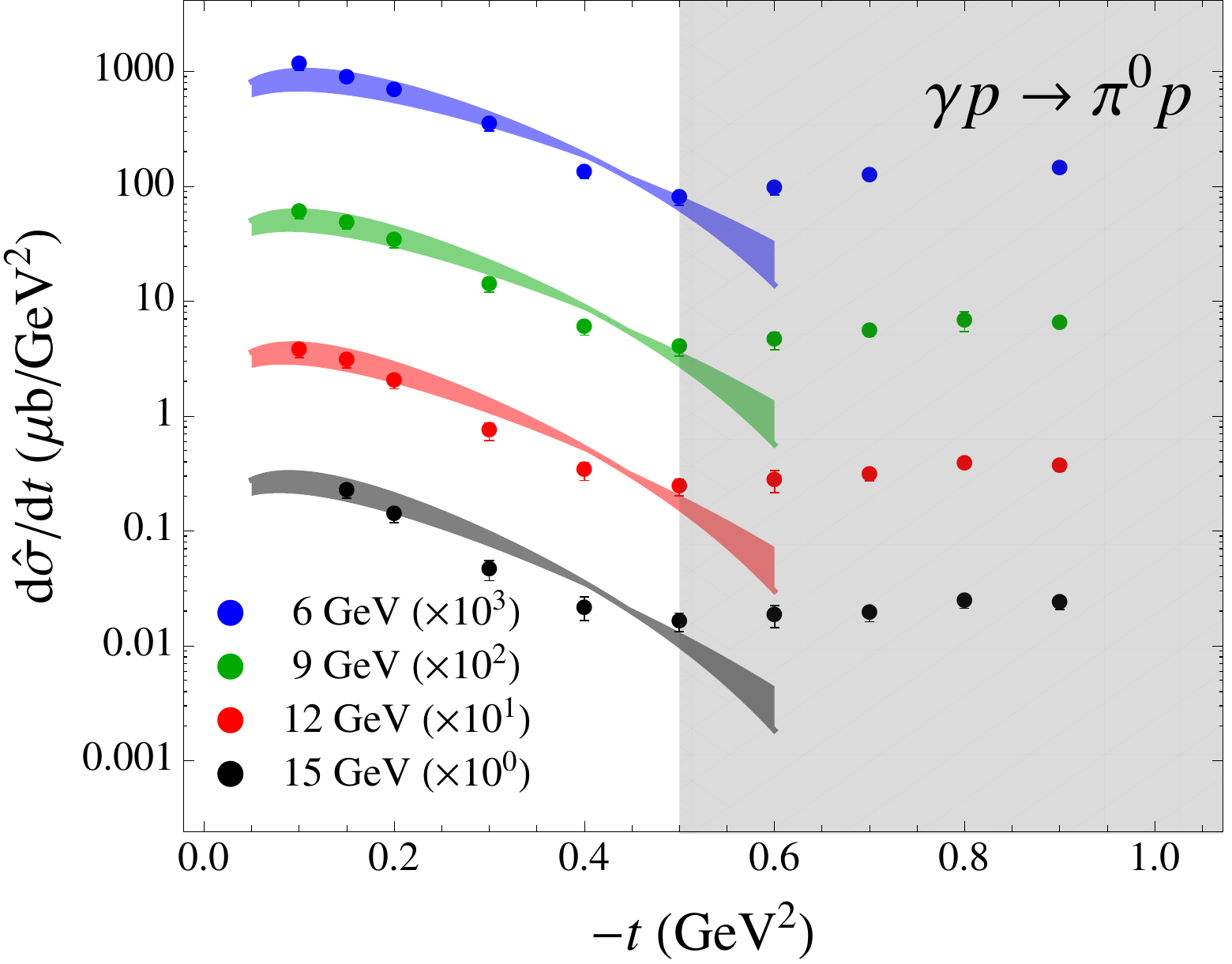}}
\: \: \: 
\subfigure[~\label{fig:Sig-Eta}]{\includegraphics[width=0.85\columnwidth]{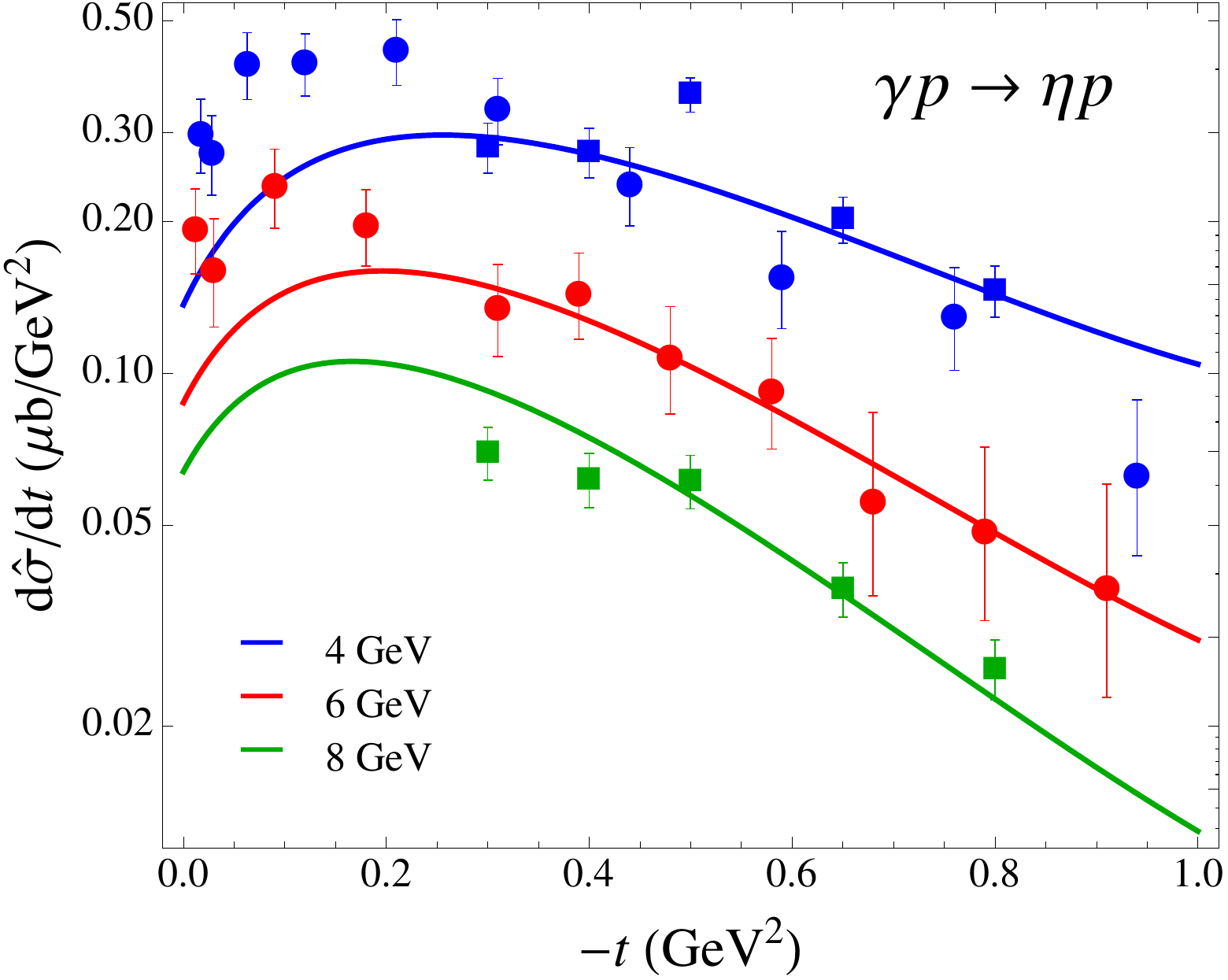}}
\end{center}
\caption{\label{fig:Sig}
Differential cross sections computed from the low energy models 
using Eq.~\eqref{eq:dsdt-bis}. 
(a)  $\pi^0$ photoproduction using SAID~\cite{Workman:2012jf}.
The prediction is restricted to $- t < 0.5\gev^2$, 
as explained in the text. 
The error band takes into account the (small) dispersion with $k$. 
The legend indicates the beam energy 
in the laboratory frame and the scaling factors. 
Data are from~\cite{Anderson:1971xh}. 
(b): $\eta$ photoproduction using \etamaid~\cite{Chiang:2001as}. 
Prediction is shown for $k=3$, as explained in the text. 
The legend indicates the beam energy in the laboratory frame.
Data are 
from~\cite{Braunschweig:1970jb} (circles) 
and~\cite{Dewire:1972kk} (squares). 
For $\pi^0$ the prediction agrees with data, 
while for $\eta$ the depletion in the forward $-t < 0.25\gev^2$ 
is a marker for an inconsistency of the low energy model.
}
\end{figure*}

The two effective residues  $\widehat\beta_{1,4}(t)$ 
are shown in Figs.~\ref{fig:Res-Pi0-1} 
and~\ref{fig:Res-Pi0-4} for $\pi^0$ 
and in Figs.~\ref{fig:Res-Eta-1} 
and~\ref{fig:Res-Eta-4} for $\eta$, respectively. 
In the case of $\pi^0$, we restrict the analysis to the 
\mbox{$0 \le -t \le 0.5\gev^2$} region, 
because of subleading Regge cut contributions 
which are known to dominate the cross section 
at higher $-t$~\cite{Mathieu:2015eia}. 
We note that the residues are fairly independent of $k$. 
Conversely, the dependence on $k$ for $\eta$ is large. 
This points to a problem in the low energy model. 
Possible reasons can be that the resonant content 
for energies less than 2 \gev is underestimated, 
or the $2-3~\gev$ resonances are relevant.
In either case the low energy model 
can be improved using these constraints.  

In Fig.~\ref{fig:Sig-Pi0} 
we predict the high energy $\pi^0$ differential 
cross section computed 
in Eq.~\eqref{eq:dsdt-bis} using the effective residues \wbi. Both the magnitude and shape of the $t$ dependence show a remarkable agreement with the data. 
The energy dependence is given by the trajectories 
in Eq.~\eqref{eq:alpha1}.
In the region of interest, the $t$ dependence is fully determined 
by the low energy amplitudes through the integral 
over the imaginary part, 
see Eq.~\eqref{eq:FESR}. 
There is a dip in the cross section data near 
$-t=0.5\gev^2$, which can be traced 
to the zero in the dominant 
$\widehat \beta_4(t)$ at $-t \simeq 0.7\gev^2$ 
in Fig.~\ref{fig:Res-Pi0-4}.
The predictions are almost independent of the moment $k$. 
The $t$ dependence is identical for moments up to $k=9$, 
and the overall normalization changes by a maximum of $20\%$. 

The predictions for $\eta$ are shown in Fig.~\ref{fig:Sig-Eta}. 
Since the \wbi computed from the low energy model 
have significant $k$ dependence, we show the cross section 
for fixed, $k=3$, which happens to have the correct overall 
normalization. 
The prediction agrees very well with data up to somewhat 
higher $-t$,  but it underestimates 
the cross section in the forward, $-t <0.25\gev^2$ region.  
This effect originates from the small value of  
$\widehat \beta_4$ in this region, 
as can be seen on Figs.~\ref{fig:Res-Eta-1} 
and~\ref{fig:Res-Eta-4}. 
It is worth noting  that the  available PWA 
models~\cite{Anisovich:2012ct,Kamano:2013iva,Ronchen:2015vfa,Chiang:2001as} 
strongly disagree in this specific $t$ region. 
In particular, in \etamaid there is a peculiar cancellation 
between isoscalar and isovector exchanges, 
which results in a smaller effective residue~\cite{Nys:2016vjz}. 
This illustrates how the implementation of our approach can 
impact on the low energy analyses. 

\textit{Conclusions.}--- We discussed a technique which uses 
analyticity to constrain low energy hadron effects 
with the high energy data. 
We have benchmarked it against meson photoproduction,
one of the main reactions to study  hadron spectroscopy.
In this specific case, we showed 
the effectiveness of the approach in 
identifying potential deficiencies in the low energy models.
We showed explicitly how the baryon spectrum determines the
seemingly unrelated meson exchanges dominating 
forward scattering at high energies, and
\textit{vice versa}. 
Experiments at Jefferson Lab are currently exploring
meson photoproduction above the baryon resonance region.
The technique presented here can be applied 
to these forthcoming data, and make a significant impact on  
baryon spectroscopy research.
The approach can be extended to other hadron reactions, 
and help  control the hadronic effects that drive 
the uncertainties in New Physics searches, 
especially in the heavy flavor sector.

\begin{acknowledgments}
\textit{Acknowledgments.}--- This material is based upon work 
supported in part by the 
U.S.~Department of Energy, Office of Science, 
Office of Nuclear Physics under contract DE-AC05-06OR23177. 
This work was also supported in part by the 
U.S.~Department of Energy under 
Grant DE-FG0287ER40365, 
National Science Foundation under 
Grants PHY-1415459 and PHY-1205019, 
the IU Collaborative Research Grant,
the Research Foundation Flanders (FWO-Flanders),
PAPIIT-DGAPA (UNAM) grant No.~IA101717,
CONACYT (Mexico) grant No.~251817, 
and Red Tem\'atica CONACYT de 
F\'{\i}sica en Altas Energ\'{\i}as (Red FAE, Mexico). 
\end{acknowledgments}

\bibliographystyle{apsrev4-1}
\bibliography{quattro}

\end{document}